\documentclass[draft]{agujournal2019}
\usepackage{url} 
\usepackage{lineno}
\usepackage{soul}
\usepackage{amsmath}
\usepackage{wasysym}
\usepackage{amssymb}
\usepackage{multirow}

\draftfalse

\journalname{JGR: Space Physics}

\begin{document}

 	\title{Time-Varying Magnetopause Reconnection during Sudden Commencement: Global MHD Simulations}
	
	\authors{J. W. B. Eggington\affil{1}, R. T. Desai\affil{1}, L. Mejnertsen\affil{1}, J. P. Chittenden\affil{2}, J. P. Eastwood\affil{1}}
	
	\affiliation{1}{Space and Atmospheric Physics Group, Blackett Laboratory, Imperial College London, London, United Kingdom}
	\affiliation{2}{Plasma Physics Group, Blackett Laboratory, Imperial College London, London, United Kingdom}
	
	\correspondingauthor{Joseph Eggington, Blackett Laboratory, Imperial College London, London SW7 2AZ, United Kingdom}{j.eggington17@imperial.ac.uk}
	
	\begin{keypoints}
	\item The dayside reconnection line responds dynamically to the compression and distortion of the magnetopause by an interplanetary shock
	\item The reconnection rate is enhanced near the shock as it propagates, and appears to be modulated by large-scale magnetopause motions
	\item The complex time-dependence of dayside coupling during SC deviates significantly from steady models of reconnection
	\end{keypoints}
	
	\begin{abstract}
	
	In response to a solar wind dynamic pressure enhancement, the compression of the magnetosphere generates strong ionospheric signatures and a sharp variation in the ground magnetic field, termed sudden commencement (SC). Whilst such compressions have also been associated with a contraction of the ionospheric polar cap due to the triggering of reconnection in the magnetotail, the effect of any changes in dayside reconnection is less clear and is a key component in fully understanding the system response. In this study we explore the time-dependent nature of dayside coupling during SC by performing global simulations using the Gorgon MHD code, and impact the magnetosphere with a series of interplanetary shocks with different parameters. We identify the location and evolution of the reconnection region in each case as the shock propagates through the magnetosphere, finding strong enhancement in the dayside reconnection rate and prompt expansion of the dayside polar cap prior to the eventual triggering of tail reconnection. This effect pervades for a variety of IMF orientations, and the reconnection rate is most enhanced for events with higher dynamic pressure. We explain this by repeating the simulations with a large explicit resistivity, showing that compression of the magnetosheath plasma near the propagating shock front allows for reconnection of much greater intensity and at different locations on the dayside magnetopause than during typical solar wind conditions. The results indicate that the dynamic behaviour of dayside coupling may render steady models of reconnection inaccurate during the onset of a severe space weather event.
	
	\end{abstract}
	
	\section*{Plain Language Summary}
	
	The Earth's magnetic field is often impacted by strong impulses of magnetised plasma ejected from the Sun. These compress the boundaries of the region shielded by the magnetic field, and generate sudden signatures in the upper atmosphere and on the ground which can have societal effects. During the compression, some mass and energy is able to penetrate this shield more effectively than usual through a process called magnetic reconnection. We perform computer simulations of several such compressions to explore the effect on reconnection in detail, quantifying its enhancement and the dependence on various parameters. We find that reconnection is intensified immediately at the point of compression and undergoes a highly time-dependent behaviour for several minutes after impact. This demonstrates that basic models of reconnection that don't account for this time-dependence may be inaccurate at the onset of severe space weather events.
	
		
	\section{Introduction}
	
	Solar wind transients such as coronal mass ejections (CMEs) and corotating interaction regions (CIRs) are responsible for a host of space weather effects, due to their interaction with the coupled magnetosphere-ionosphere system. These can contain large out-of-ecliptic magnetic field components, and carry interplanetary (IP) shocks at their leading edge. The sudden increase in solar wind density and/or velocity is characterised as a dynamic pressure enhancement (DPE), which results in strong compression of the magnetopause. The ground signature of this compression is a sharp, bipolar variation in the horizontal magnetic field (e.g. \citeA{Smith2019}). This is termed the geomagnetic sudden commencement (SC), which if it develops into a geomagnetic storm is known as storm sudden commencement (SSC), or a sudden impulse (SI) if it does not \cite{Araki2013}. SSCs are increasingly recognised as a space weather threat to power systems, as they can induce particularly large GICs \cite{Eastwood2018}. 
	
	IP shocks represent an extreme type of DPE, and their effect on the magnetosphere has been widely studied in global magnetohydrodynamic (MHD) simulations. Once a fast forward IP shock front reaches the Earth's bow shock it is decelerated within the dense magnetosheath, forming a curved front which then compresses the magnetopause \cite{Samsonov2006}. Waves are consequently launched into the magnetosphere, which have been shown to propagate faster than the shock front in the solar wind, especially on the nightside as they travel into the magnetotail \cite{Andreeova2008}. Simulations have further shown that the reflection of fast mode waves off the simulation inner boundary (representing reflection by the ionosphere) acts to decelerate the compression of the magnetopause and bow shock, and which further reflects back towards the inner boundary (\citeA{Samsonov2007}, \citeA{Yu2011}). Similar simulations have been performed for dynamic pressure decreases \cite{Ozturk2019}, e.g. due to reverse shocks, which have the opposite effect of causing expansion of the boundaries. Theoretical studies predict that oscillations in the magnetopause position occur when it is displaced \cite{Freeman1995}, as has been shown on the scale of several Earth radii in response to IP shocks in recent MHD simulations \cite{Desai2021a}. Any such oscillations may also modulate the local reconnection rate and flux transfer event (FTE) generation along the dayside magnetopause.
	
	It is the initial fast mode shock wave, transmitted through the magnetopause, which first triggers the ground signatures defining SC. Two distinct phases are seen, called the preliminary impulse (PI) and main impulse (MI) \cite{Araki2013}, associated with intense field-aligned current (FAC) signatures in the ionosphere. Global MHD simulations have been used to explore the response of the FAC to a sudden pressure pulse in the solar wind in great detail (e.g. \citeA{Slinker1999}, \citeA{Keller2002}, \citeA{Fujita2003a}, \citeA{Fujita2003b}). In a study of shock impact during northward interplanetary magnetic field (IMF), \citeA{Samsonov2010} described these currents in terms of transient FAC systems which then decay to whatever conditions arise due to the post-shock solar wind driving. This included a transient Region 1-like system resembling that typically associated with southward IMF, and which has been attributed specifically to equatorial flow vortices in the outer magnetosphere generated near the shock front (\citeA{Samsonov2013}, \citeA{Tian2016}, \citeA{Zhao2016}, \citeA{Kim2017}). A two-phased FAC response has also been seen in simulations using southward IMF \cite{Yu2009b}, for which any convection-driven FAC signatures may be more intense given the stronger reconnection dynamo. 
	
	The response of the ionospheric polar cap to a change in driving conditions depends on the extent to which dayside and nightside reconnection are enhanced. Using the OpenG\nolinebreak[5]GCM global MHD code, \citeA{Oliveira2014} studied the geoeffectiveness of frontal and inclined shocks (defined with respect to the Sun-Earth line). The dayside ionosphere showed strong general FAC signatures due to the compression, whilst frontal shocks were more geoeffective than inclined shocks as they triggered the closure of nightside magnetotail flux, i.e. substorms. Similar conclusions have been made from observations of the FAC response \cite{Shi2019}. Such nightside flux closure is associated with contraction of the polar cap, and this has been widely observed to occur following DPEs (e.g. \citeA{Milan2004}, \citeA{Boudouridis2005}, \citeA{Hubert2006}, \citeA{Boudouridis2008}, \citeA{Hubert2009}).
	
	Nonetheless, dayside reconnection can also be enhanced following the arrival of a DPE, as has been shown in observations (e.g. \citeA{Boudouridis2007}) and simulations (e.g. \citeA{Ober2007}). Furthermore, \citeA{Samsonov2010} associated transient FACs in their simulation of shock propagation with enhanced lobe reconnection. Hence both dayside and nightside reconnection must compete to determine the change in polar cap and auroral oval shape and size in the short time after IP shock impact, which will be complicated by time-delays in the response. Based on a large number of events \citeA{Boudouridis2011} showed prompt dayside convection in response to DPEs under southward IMF, which was slightly delayed and weaker but longer lasting for northward IMF, and with nightside responses delayed by $\sim$ 10-15 min. 
	
	In order to separate the dayside and nightside reconnection rate responses to a DPE, \citeA{Connor2014} performed simulations using OpenGGCM and calculated the rates via the ionospheric potential. For southward IMF they showed a clear increase of the dayside rate after the Alfv\'{e}n wave transit time into the ionosphere, with the nightside rate responding several minutes later. They attributed this increase to a spike in the compressed magnetosheath field due to the DPE, and hence expansion followed by eventual contraction of the polar cap. Recently, \citeA{Boudouridis2021} performed a similar case study comparing observations of the polar cap response to OpenGGCM simulations. Whilst the simulation predicted an initial noon-to-afternoon expansion of the polar cap, the observations showed a contraction at all magnetic local times (MLT) for which there was sufficient data, even on the dayside. Nonetheless, the observations showed that the dayside rate did in fact promptly increase, which according to the expanding/contracting polar cap (ECPC) paradigm \cite{Cowley1992} should result in some expansion. Unless some effect is preventing this from occurring, any expansion must therefore be smaller and/or harder to detect than the subsequent contraction, and was possibly overestimated in the simulation.
	
	Overall it is clear that dynamic pressure enhancements, particularly IP shocks, can result in immediately enhanced dayside reconnection which would contribute to any resulting space weather impacts. However, the extent to which any enhancement occurs, its dependence on the specific driving conditions and the local time-dependent response along the reconnection X-line remain unclear. These issues are important for assessing the accuracy of empirical solar wind coupling functions which would simply predict a step change in the total reconnection rate in response to a shock. There are further implications for studies of reconnection at the magnetopause. The compression of the boundary will influence where reconnection occurs given that the magnetic separator (the 3-D X-line) should evolve with the deformed surface, which may alter the local reconnection rate due to changes in the field orientation and local plasma conditions. Whether the separator can evolve dynamically over a given timescale is very difficult to verify with in-situ measurements; observations of reconnection X-lines by MMS have shown stationarity of the X-line over several minutes, but these were for steady IMF conditions \cite{Fuselier2019}. 
	
	Global MHD simulations therefore provide an ideal means to investigate these issues, given their ability to capture the dynamical changes in the global system in the short time following impact. In this study, we simulate the response of the separator and the associated ionospheric coupling due to a variety of shocks impacting the magnetosphere using the Gorgon MHD code. By simulating a series of different clock angles, dipole tilt angles and solar wind dynamic pressures we can examine how these parameters control the response of the dayside coupling. Furthermore, by repeating our simulations with an artificial resistivity, we can investigate in close detail how the reconnection rate is altered locally as the shock propagates along the magnetopause.		
	
	
	\section{Simulation Set-up}
	
	The Gorgon MHD code solves the semi-conservative resistive MHD equations on a uniform, regular Eulerian cartesian grid to second order, with advection terms handled with the third-order Van Leer advection scheme (\citeA{VanLeer1977}, \citeA{Ciardi2007}, \citeA{Mejnertsen2018a}). We employ a high-resolution grid of spacing 0.25 $R_E$ everywhere in the simulation domain, as discussed in \citeA{Desai2021b}, spanning $X$ = (-30, 90) $R_E$, $Y$ = (-40, 40) $R_E$, $Z$ = (-40, 40) $R_E$. These coordinates are related to the Geocentric Solar Magnetospheric system (GSM) by ($X$, $Y$, $Z$) = ($-X_{GSM}$, $-Y_{GSM}$, $Z_{GSM}$). The inner boundary is placed at 3 $R_E$ and has plasma parameters of density $n$ = 370 cm$^{-3}$ and ion/electron temperature $T_{i,e}$ = 0.1 eV \cite{Eggington2020}. The FAC is mapped from a fixed radius along dipole field lines to a thin-shell ionosphere model with a uniform Pedersen conductance of 10 mho and zero Hall conductance \cite{Eggington2018}. The ionospheric potential is updated every 5 s, and mapped back out to the inner boundary where it is used to calculate the E$\times$B-drift as an inner boundary condition. 
	
	Gorgon utilises a Wilkins artificial viscosity which confines shocks to a few grid cells and removes spurious oscillations \cite{Wilkins1980}. This is combined with a Christensen flux-limited viscosity which applies a first order correction and slope-limiting to the velocity jumps used to compute the artificial viscosity. The velocity is assumed to vary linearly over the cell, lowering the viscosity in regions with smooth solutions, whilst the slope-limiting (similar to that in the Van Leer advection scheme) reduces the solution to the Wilkins viscosity in regions with sharp discontinuities hence providing shock dissipation where it is needed \cite{Benson1992}. Since the width of any shocks in the simulation is then naturally determined by the grid resolution, our use of a uniform 0.25 $R_E$ grid spacing results in a relatively high resolution in the solar wind, allowing for more detailed analysis of time-variations in the magnetopause response. Note that the code also applies a Boris correction to limit the Alfv\`{e}n speed, with an artificial speed of light of $10^7$ ms$^{-1}$ \cite{Boris1970}. This introduces some model-dependence in the propagation times of disturbances through the system. 	
	
	The magnetosphere is initialised for 2 hours in each of our runs, with solar wind conditions of $n$ = 5 cm$^{-3}$, $v_x$ = 400 kms$^{-1}$, $T_{i,e}$ = 5 eV and $B$ = 2 nT. A shock is then injected from the sunward edge of the box by introducing a jump in solar wind parameters that satisfies the Rankine-Hugoniot (RH) jump conditions for mass and momentum conservation \cite{Priest1982}. The post-shock solar wind parameters are then kept the same for the remainder ($\sim10$ min) of the simulation. We simulate five different shocks in total, with jump conditions based on those used in \citeA{Desai2021a}. These all yield shock speeds of 200 kms$^{-1}$ in the post-shock solar wind frame, representative of shocks typically observed at 1 AU \cite{Berdichevsky2000} and should therefore be broadly representative of impact by a CME-driven shock resulting in SI/SSC.	
	
	Firstly to investigate the dependence of the reconnection response on the IMF orientation, Shocks 1, 2 and 3 have three different IMF clock angles ($\theta_{IMF}$ = $\tan^{-1}(B_{y,GSM}/B_{z,GSM})$) of 180$^\circ$ (due southward), 135$^\circ$ (equally southward and duskward) and 90$^\circ$ (due duskward) respectively, assuming no dipole tilt. According to the coplanarity theorem, the parallel field orientation is unchanged through the shock and so the magnetosphere is initialised separately for each case of $\theta_{IMF}$. Shock 4 is the same as Shock 1 except that the dipole tilt angle $\mu$ = 30$^\circ$, representing an event occurring during Northern summer to explore any seasonal dependence. Finally, Shock 5 is a stronger version of Shock 1 with a greater jump in density and velocity; the four-fold increase in density is the upper limit allowed by the RH conditions for the $\gamma$ = 5/3 case. In each case the jump in IMF magnitude is kept the same so as to isolate the effect of other parameters. The full set of different simulation conditions is shown in Table \ref{tab:shock_jump_conds}.
	
	\begin{table}[ht!]
		\centering
		\begin{tabular}{c|cccccc}	
			Shock & $\theta_{IMF}$ / $^\circ$ & $\mu$ / $^\circ$ & $n$ / cm$^{-3}$ & $v_x$ / kms$^{-1}$ & $T_{i,e}$ / eV & $B$ / nT \\ \hline
			1 & 180 & 0 & 10 & 600 & 417 & 4 \\ 
		    2 & 135 & 0 & 10 & 600 & 417 & 4 \\
			3 & 90  & 0 & 10 & 600 & 417 & 4 \\
			4 & 180 & 30 & 10 & 600 & 417 & 4 \\
			5 & 180 & 0 & 20 & 1000 & 1250 & 4 \\ 
		\end{tabular}
		\caption{Different conditions used for each simulated shock, showing the IMF clock angle, dipole tilt angle, and post-shock solar wind parameters. The pre-shock solar wind parameters are the same in each case and are as follows: $n$ = 5 cm$^{-3}$, $v_x$ = 400 kms$^{-1}$, $T_{i,e}$ = 5 eV and $B$ = 2 nT.}
		\label{tab:shock_jump_conds}
	\end{table}
		
	Before proceeding we note that in order to avoid any effects of IMF $B_x$ on the location of reconnection on the magnetopause, we set $B_x = 0$ in each case. This requires assuming purely perpendicular shocks, across which the quantity $1/2v^2 + c_s^2/(\gamma-1) + v_A^2$ (where $c_s$ and $v_A$ are the sound speed and Alfv\'{e}n speed respectively) is not strictly conserved. Combined with the effects of discretisation on the grid, the result is a breakdown of the discontinuity into a two-step profile: a steep transient shock front of small but finite width followed by a slight undershoot, which then tends to steady conditions corresponding to the values shown in the table. Specifically, the conditions directly at the shock fronts correspond to a four-fold increase in $n$ and $B$, i.e. to 20 cm$^{-3}$ and 8 nT respectively. Therefore we caution the reader that the actual jumps at the bow shock are not that of a simple step change, and where this is important it is considered in our analysis. However, the shock fronts have a transit time of only $\sim$ 10-20 s at a given point on the magnetopause, such that on timescales of several minutes the global response will approximate that of a simple step change and any differences are generally short-lived.

	The shocks are simulated first with no resistivity included, such that the magnetosphere is collisionless and the compression of the magnetopause is physically representative. Reconnection is thus mediated by numerical diffusion in these five runs, and it is these we use for the majority of our analysis. For the analysis in section \ref{sec:resistive_runs} we then repeat Shocks 1, 3 and 5 while including an explicit resistivity of $\eta/\mu_0 = 5\times10^{10}$ m$^2$s$^{-1}$, which allows us to directly calculate and closely examine the evolution in the reconnection electric field along the magnetic separator, and acts to smooth-out current sheets for ease of analysis. This value for $\eta$ is also sufficiently large to dominate over numerical diffusion, reducing any model-dependence in the results and hence improving the reproducibility of our findings; comparable values have been used in other studies which explored reconnection in global simulations (e.g. \citeA{Glocer2016}). 

	
	\section{Global Topology during Shock Propagation}
	
	\subsection{Magnetospheric Response}
	
\begin{figure}[h!]
		\centering
		\includegraphics[width=\linewidth]{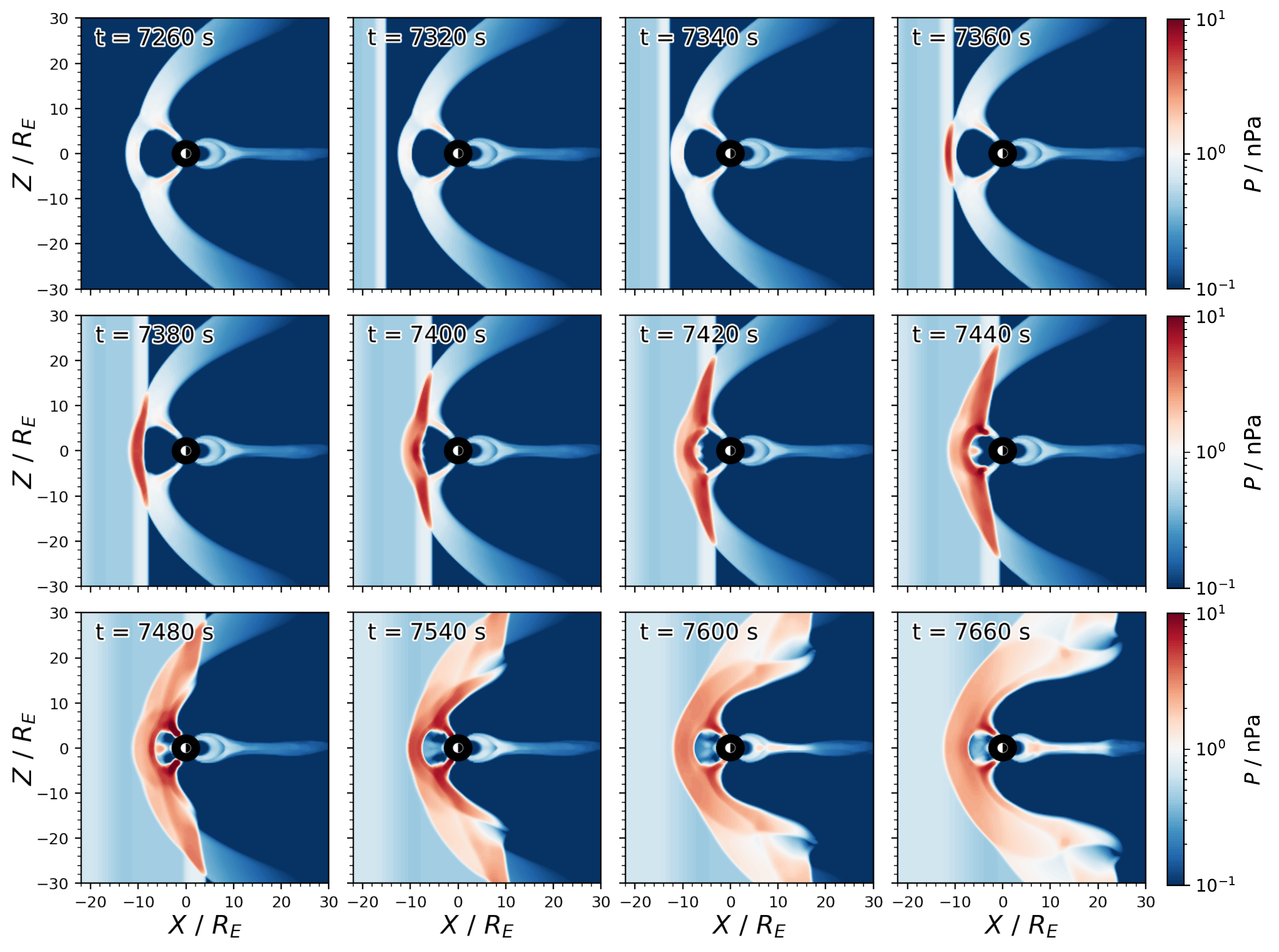}
		\caption{Slices in the noon-midnight plane for Shock 1, showing the thermal pressure $P$ over time as the shock propagates through the magnetosphere.}
		\label{fig:pressure_shock} 
	\end{figure}

	Figure \ref{fig:pressure_shock} shows the evolution of the thermal pressure during the period of propagation of Shock 1. The shock can be seen arriving in from the left-hand edge, making contact with the bow shock at around 7340 s and the magnetopause at $\sim10$ $R_E$ at around 7360 s. Since the solar wind conditions prior to contact are time-constant, the system is initially in a quasi-steady configuration. By 7380 s the jump in dynamic pressure has passed the bow shock and compressed the magnetopause, and the shock begins to advance towards the flanks, distorting the magnetopause surface at the point of contact. By 7480 s the shock has passed the terminator plane, and the magnetopause has approached its minimum stand-off distance.  
	
	A fast wave is seen propagating throughout the dayside magnetosphere from 7420 s onwards as a pressure front which then reflects off the inner boundary. By 7540 s the magnetopause stand-off distance has returned to $\sim8$ $R_E$, after which it will undergo some further lower-amplitude oscillations until it relaxes to balance the post-shock driving conditions \cite{Desai2021a}. This complex overall motion will drastically change the magnetosheath flow in the frame of the magnetopause, and thus should modulate the coupling with the solar wind.
	
	On the nightside, the plasma sheet pressure is enhanced from $\sim$ 7600 s, roughly 4 min after the initial impact. Whilst we do not explicitly investigate the nightside reconnection response in the present study, we would expect any enhancement to begin around this time for Shocks 1-4, and slightly earlier for the faster-propagating Shock 5. However, the distance (and hence delay) at which any subsequent nightside reconnection is triggered will to some extent depend on the grid resolution in the tail and the numerics of the model, as well as being sensitive to the preceding driving conditions (e.g. the amount of open flux loaded in the tail). It is therefore difficult to generalise the effect on tail reconnection and the characteristic timescales over which it is enhanced purely based on the observed enhancement of the plasma sheet pressure. In any case, it is apparent that there will be an initial period of $\gtrsim$ 5 min prior to the nightside response where any enhancement in dayside reconnection will increase the amount of open flux in the system. This should thus expand the polar cap in the absence of enhanced nightside reconnection, and we focus our analysis on this time period.
	
	\subsection{Ionospheric Response}
	
	The ionospheric signatures of enhanced dayside reconnection are the growth of open flux in the polar cap and the excitation of convective flows. This should manifest initially as an expansion of the dayside open-closed field line boundary (OCB), and strong associated Region 1 FACs. Whilst the ionospheric response to SI/SSC has been widely studied in the literature, we will focus here on only the first several minutes after onset. Once again we only examine the case of Shock 1, to obtain a general overview of the sequence of events in the system response. The timescales of this will be broadly consistent with Shocks 2-4 and be slower than those for Shock 5. 
	
\begin{figure}[h!]
		\centering
		\includegraphics[width=\linewidth]{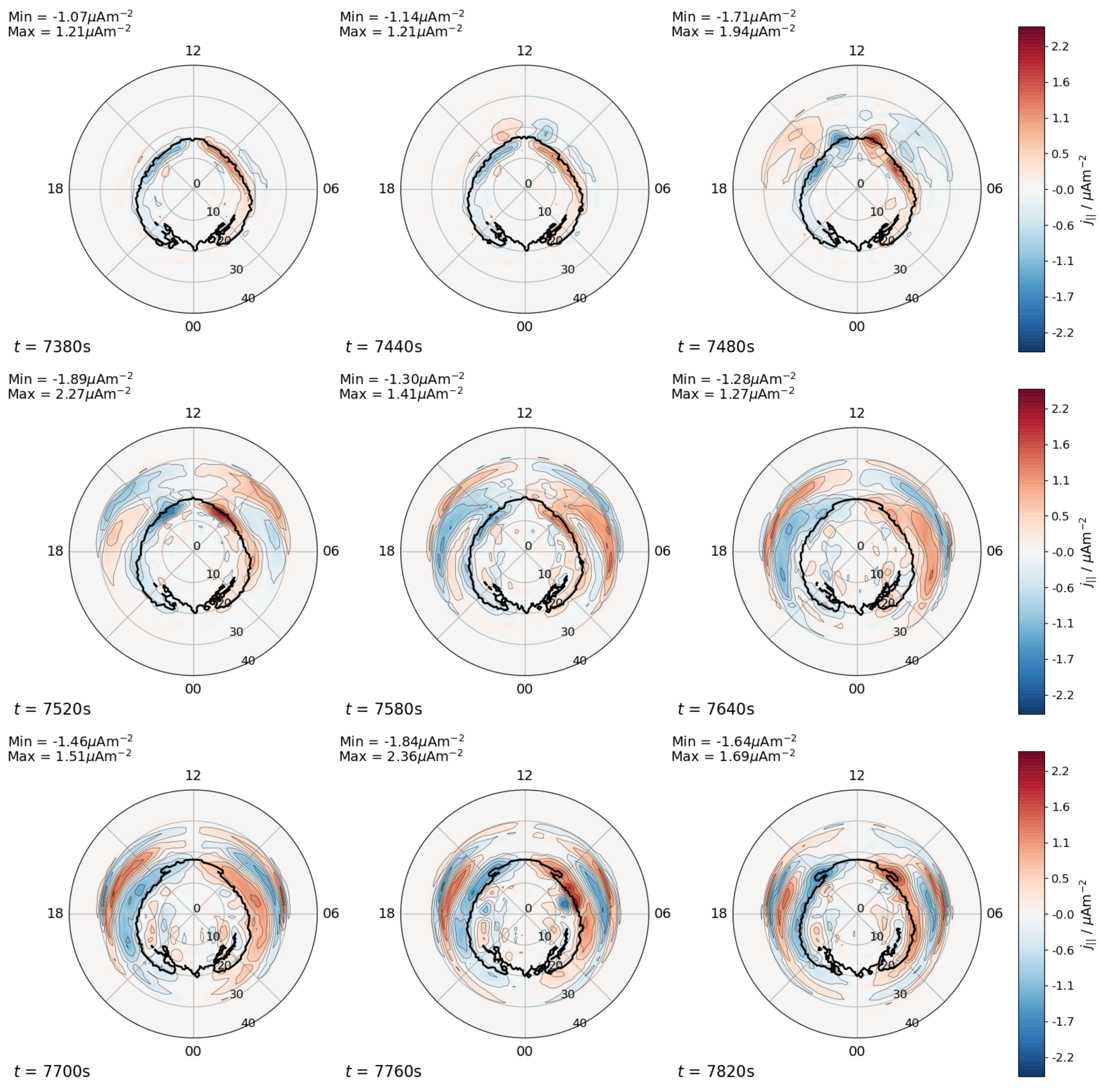}
		\caption{Ionospheric field-aligned current $j_\parallel$ in the Northern hemisphere during the propagation of Shock 1. The open-closed field line boundary is indicated by the black line each case.}
		\label{fig:ionoshock}
	\end{figure}

	Figure \ref{fig:ionoshock} shows the Northern ionospheric FAC during the propagation of Shock 1. The OCB is indicated by the black line, and found by sampling the magnetic connectivity at the simulation inner boundary. The first signatures are seen at 7440 s, $\sim$ 80 s after the shock arrival. At pre-noon and post-noon, the current profile becomes bipolar, with oppositely-directed FAC appearing at lower latitudes just outside of the polar cap region. These are in the same sense at the expected additional FAC signatures during the PI phase (e.g. \citeA{Fujita2003a}). 
	
	The higher latitude FACs grow in intensity around noon, reaching $\sim2$ $\mu$Am$^{-2}$ in magnitude at 7520 s, at which point the first compressional signatures move into the nightside. An additional pair of oppositely-directed currents appear equatorward of these and eventually merge into the original Region 1 FACs by 7700 s. The result is a much stronger Region 1 system lying at about 5-10$^\circ$ lower in latitude than for the pre-shock conditions, which remains steady unlike the transient current systems associated with the compressional wave at the front of the shock. Further examples of the latter appear equatorward even as late as 7820 s, likely resulting from additional reflected pressure waves in the inner magnetosphere, as suggested in other simulations in which similar signatures were seen \cite{Yu2011}. Note however that the intensity and duration of these poleward-moving FACs, which originate near the equator of the inner boundary, will be sensitive to the boundary conditions in the model and hence may be exaggerated in the simulation.
	
	The signatures in the OCB are hard to distinguish at first, though its morphology is slightly changed either side of noon at 7480 s coincident with the enhancement in the Region 1 FAC. Clearer effects are seen between 7580-7640 s, where it begins to expand in MLT around noon and move down to slightly lower latitudes. Between 7700-7820 s this expansion of the OCB proceeds towards dawn and dusk, where regions of open flux associated with strong Region 1 FAC spread into the morning and evening sectors. This behaviour is consistent with the expectations of enhanced dayside reconnection generating stronger convection. No clear nightside OCB signatures are seen until after the final time shown here. This is consistent with the sequence of events in Figure \ref{fig:pressure_shock}, since we only expect enhanced nightside reconnection from around 7600 s onwards, and with a delay in the FAC at least comparable to the 100 s delay on the dayside.
	
	Overall whilst the immediate ionospheric response is dominated by the effects of the compression of the magnetopause, the underlying reconnection-driven FAC signatures continue to manifest and form a clearly recognisable, expanded Region 1 system after about 5 min. The expansion of the dayside polar cap occurs over this entire period, consistent with OpenGGCM simulations of the initial polar cap response to a DPE \cite{Boudouridis2021}. Whilst we expect differences in the polar cap response for the other simulated shocks, examining these is outside the scope of this study as our primary focus in comparing the effect on dayside reconnection.
	
	
	\section{Magnetopause Reconnection Impact}
	
	\subsection{Separator Evolution} 
	
	To study the evolution of dayside reconnection, we trace-out the location of the magnetic separator (the 3-D X-line) on the magnetopause. This line demarcates different magnetic domains, and is controlled by the orientation of the IMF and magnetospheric field. For a 180$^\circ$ clock angle, i.e. purely southward IMF (as with Shocks 1 and 5), the separator lies essentially along the equatorial plane. Due to the different field orientations this will not be the case for Shocks 2-4. Specifically, a change in clock angle causes the separator to rotate about the GSM $X$-axis, such that it is aligned with the $X$-$Z$ plane for due northward IMF \cite{Komar2013}. More detailed discussion of the effect of IMF orientation on the location of magnetopause reconnection can be found in the literature (e.g. \citeA{Yeh1976}, \citeA{Crooker1979}, \citeA{Pudovkin1985}, \citeA{Alexeev1998}). Meanwhile, as shown by previous studies (e.g. \citeA{Hoilijoki2014}, \citeA{Eggington2020}), an increase in the dipole tilt angle causes the separator to shift from the subsolar magnetopause and decrease in length. The tilt and clock angles also influence the configuration of the magnetotail current sheet (e.g. \citeA{Xiao2016}) and hence may additionally affect the subsequent shock impact on tail reconnection. 
	
	For Shocks 1 and 5 the separator location can be approximated simply by finding the locus of points where $B_z$ = 0, which corresponds to the magnetopause. Though it strictly may deviate somewhat in the $Z$-direction, we are not initially concerned about sampling the reconnection rate locally along the separator and therefore there is little benefit in attempting to trace it out in fine detail, which is more computationally expensive. For Shocks 2-4, such an approximation is not possible and we must trace the separators in full, employing the same separator tracing algorithm as described in detail by \citeA{Eggington2020}, and similar to that of \citeA{Komar2013}.
	
\begin{figure}[h!]
		\centering
		\includegraphics[width=\linewidth]{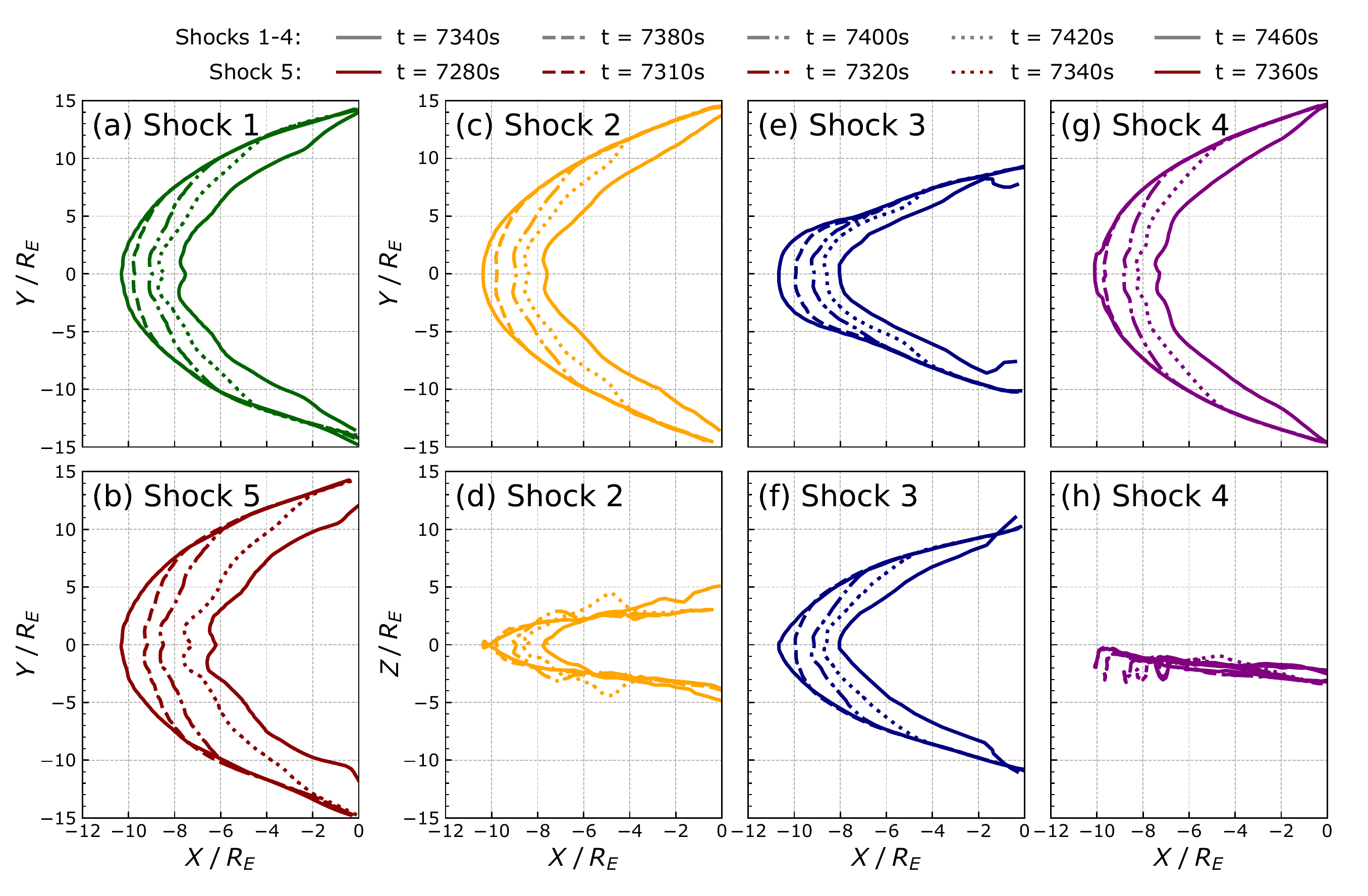}
		\caption{Evolution of the dayside magnetic separator for the simulated shocks, shown in the $X$-$Y$ plane (a, b, c, e, g) in each case and also the $X$-$Z$ plane (d, f, h) for Shocks 2, 3 and 4. Note the timestamps shown for Shock 5 in panel (b) are different to the other panels due to its faster propagation speed.}
		\label{fig:Separators_2D} 
	\end{figure}

	Topologically, a separator connects a pair of magnetic null points (where $|\mathbf{B}|$ = 0), which on the magnetopause should lie near the terminator plane in opposite hemispheres. These have therefore been used in previous simulations as the start point for tracing the dayside separator (e.g. \citeA{Komar2013}, \citeA{Glocer2016}, \citeA{Eggington2020}). However, due to the compression of the magnetopause and resulting changes in the magnetic topology, this becomes more complicated. Instead we sample the magnetic connectivity on the terminator plane to find the convergence point, i.e. intersection of the separator, and trace from there. In order to track the separator evolution during the magnetopause compression phase we choose only a few sample timesteps: these are prior to the shock impact, during the propagation along the dayside, and just before magnetopause reaches its minimum stand-off distance. The evolution of the dayside separators for these timesteps is shown in Figure \ref{fig:Separators_2D}.
	
	The subsolar portion of the separators which first makes contact with the shock is compressed immediately and continues to move inwards even after the shock front has proceeded towards the flanks. The displacement is greatest for Shock 5, due to its larger jump in solar wind dynamic pressure. In all cases there is some indentation in the subsolar magnetopause, most apparent at the later timesteps in the $X$-$Y$ plane, which indicates erosion of dayside flux due to enhanced reconnection. This effect is most severe for Shocks 1 and 5, which is to be expected as due southward IMF is most favourable for reconnection. Since the magnetopause is in motion during the entire compression phase (with an average subsolar speed of $\sim 130$ kms$^{-1}$ in the case of Shocks 1-4) this will alter the magnetosheath electric field within the frame of the separator. We therefore expect a non-linear response in the reconnection rate. 
	
\begin{figure}[h!]
		\centering
		\includegraphics[width=\linewidth]{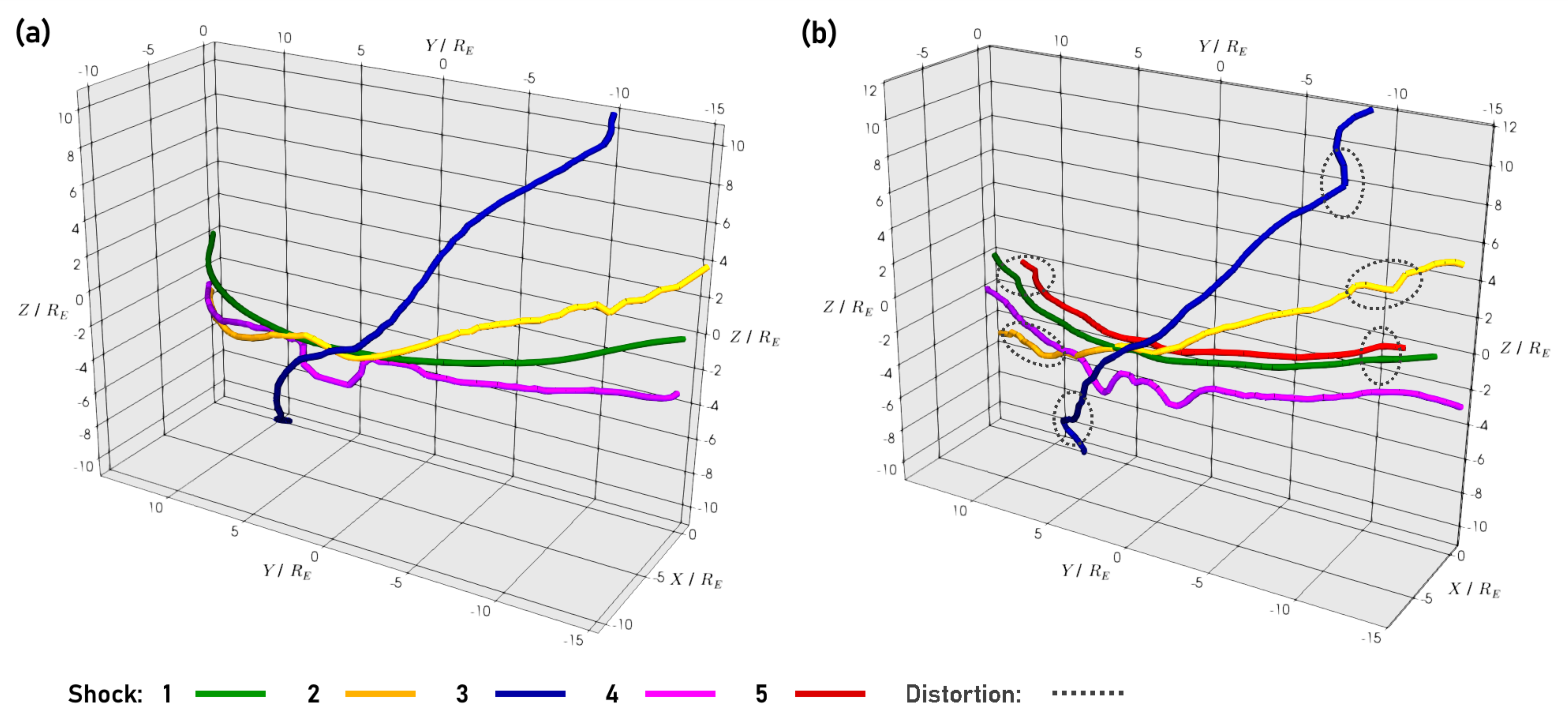}
		\caption{3-D perspective of the dayside magnetic separator for each shock, shown at the (a) the start and (b) the end of the initial compression phase, i.e. $t = 7340$ s and $7460$ s for Shocks 1-4 and $t = 7280$ s and $7360$ s for Shock 5. Regions where there is visible distortion of the separator due to the presence of the shock front are indicated.}
		\label{fig:Separators_3D} 
	\end{figure}

	A key feature of the evolution is that downstream segments of the separator remain in their initial configuration until they have been processed by the shock. This is particularly noticeable for Shocks 2 and 3, since they are inclined away from the equatorial plane; for the former, the separator is lifted in the $Z$-direction at the shock front at 7400 s and 7420 s, which is a significant deviation from any steady configuration. Figure \ref{fig:Separators_3D} shows a 3-D perspective of the separators at the initial and final timesteps shown in Figure \ref{fig:Separators_2D}, with the distortion of the separator indicated. This incoherence in the separator response will be associated with a strong magnetopause current at the point of compression, thus altering the local reconnection electric field and complicating the existing clock angle-dependence in the total reconnection rate. 
		
	Whilst the distortion by the shock is not as visibly sharp in the case of Shock 4, more complex behaviour is seen around noon. Due to the dipole tilt the separator is shifted southward from that of Shock 1, and branches away from the subsolar point indicating the occurrence of an FTE. Note that as in \citeA{Eggington2020} we only attempt to trace-out one such branch of the separator, though others may exist around the FTE. The southward displacement from the equatorial plane remains after the compression, though its geometry around noon varies significantly, suggesting that complex magnetic field structures can survive and continue to evolve on the compressed magnetopause surface. 

	\subsection{Dayside Reconnection Enhancement}
	
	To understand how the separator evolution relates to dayside coupling we now attempt to quantify both the local and total reconnection rate; the former is given by the electric field parallel to the separator, whereas the latter is given by the line integral of this quantity over the length of the separator. In the absence of parallel electric fields and in steady-state, this quantity maps down as the ionospheric cross-polar cap potential (CPCP) \cite{Hesse1997}. However due to the strong compression of the magnetosphere, the system is far from steady-state. This means that directly relating ionospheric convection to dayside reconnection can be difficult, especially given any time-delays in the ionospheric response. 
	
	For a detailed analysis of the dayside reconnection rate over the short time interval of interest, we therefore avoid relying on ionospheric signatures (as done for a longer time period by \citeA{Connor2014} and \citeA{Boudouridis2021}) and focus on the immediate change in the amount of open flux in the system. Since the magnetosphere is initially quasi-steady prior to the shock arrival, the dayside and nightside reconnection rates are largely in balance and the dayside rate $\Phi_D$ and polar cap flux content $F_{PC}$ can be taken as roughly constant. For the value at some general time $t$ after the time of impact $t_0$ -- and prior to the compression of the magnetotail current sheet after $5$ min -- the dayside rate can then be approximated as
	
	\begin{linenomath*}
		\begin{equation}
			\centering \Phi_D(t) \approx \Phi_{D}(t_0) + \frac{dF_{PC}(t)}{dt}.
			\label{eq:shock_reconn_rate}
		\end{equation} 
	\end{linenomath*}

\begin{figure}[h!]
		\centering
		\includegraphics[width=\linewidth]{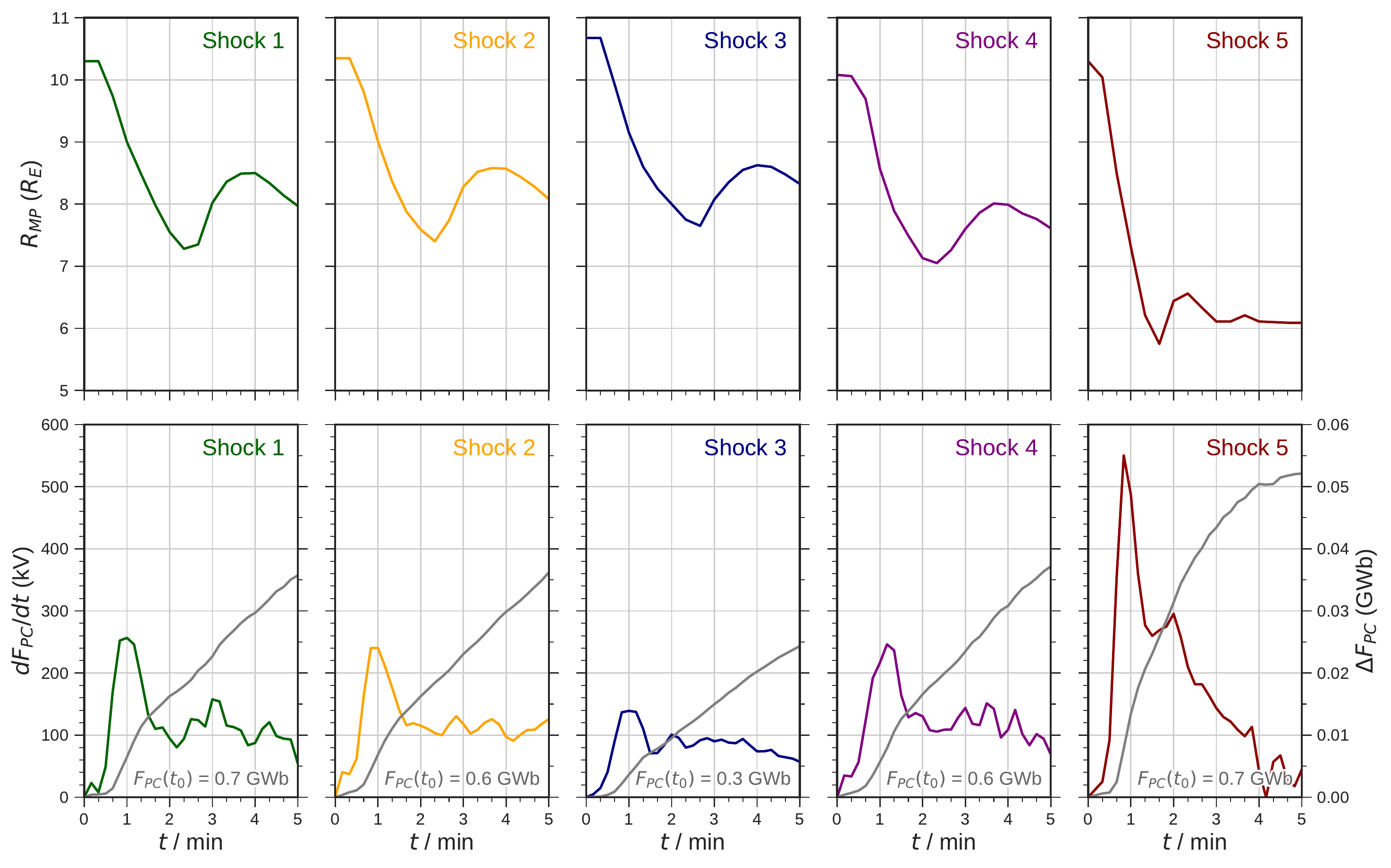}
		\caption{Magnetopause stand-off distance $R_{MP}$ (top row) over time along with the change $\Delta F_{PC}$ (bottom row, grey lines) and rate of change $dF_{PC}/dt$ (bottom row, coloured lines) of open flux in the Northern polar cap for each shock, shown for a period of 5 min of propagation after impact. The initial open flux content prior to impact in each case is indicated.}
		\label{fig:flux_change_vs_time} 
	\end{figure}

	The second term represents the direct impact of the IP shock on $\Phi_D$ regardless of the preceding strength of driving. Figure \ref{fig:flux_change_vs_time} shows the magnetopause stand-off distance, growth of open flux and value of $dF_{PC}/dt$ in the Northern hemisphere over the first 5 min of shock propagation for each event. For Shocks 1, 2, 4 and 5 the stand-off distance is identified as the point along the subsolar line where the sign of $B_z$ reverses; for Shock 3 where the IMF $B_z$ = 0 we instead find where the magnetic connectivity changes from closed magnetospheric field to solar wind field. The open flux content is calculated by sampling magnetic connectivity at the outer boundaries of the simulation box (see \citeA{Eggington2020}), sampled every 10 s. Note the time $t$ is zeroed at 7340 s for Shocks 1-4 and at 7280 s for Shock 5 (which propagates more quickly). 
	
	For each shock there is an initial period of compression to a minimum stand-off distance, followed by an expansion (before another compression) due the reflection of a pressure wave off the inner boundary \cite{Samsonov2007}. For Shock 5 this occurs more rapidly due to the greater dynamic pressure, and further lower-amplitude oscillations are seen as described by \citeA{Desai2021a}. In each case the arrival of the shock results in a sharp increase in the dayside reconnection rate, which then drops from its peak as the compression of the magnetopause begins to slow. The size of this peak will to some extent depend on the overshoot at the leading edge of the shock, though in internal investigations where we injected the shock from much closer to the bow shock (so as to minimise any overshoot) we obtained much the same response. Therefore this initial spike must result in part from the sudden pile-up of shocked magnetosheath plasma and magnetic field, in agreement with findings by previous studies (\citeA{Connor2014}, \citeA{Boudouridis2021}). 
	
	Examining Shocks 1-3, the peak increase in reconnection rate shows a much weaker clock angle-dependence than the $\sim\sin^4(\theta_{IMF}/2)$ used in various coupling functions (e.g. \citeA{Milan2012} and references therein), many of which assume a linear dependence on velocity and would predict values for Shock 1 to be around four times that of Shock 3. We note however that other simulation studies have also found a weaker clock angle scaling of dayside coupling than predicted by some empirical and theoretical formulae (e.g. \citeA{Wang2014}, \citeA{Komar2016}), so this is likely not a unique effect of the compression. Similarly, the introduction of a dipole tilt for Shock 4 makes little difference to the peak rate. Nonetheless the fact that a strong enhancement is seen for all orientations and persists for some time after impact shows that the jump in dynamic pressure has a significant control over the strength of coupling, which is not greatly limited by the clock or tilt angle. 
	
	This is clear from Shock 5, which has a substantially greater peak $dF_{PC}/dt$ than Shock 1 with the sole distinguishing factor being the dynamic pressure jump. However whilst the speed of the final post-shock solar wind differs by a factor of 1.67, the ratio of the peak rates is greater at 2.14. Since the final post-shock density in Shock 5 is greater than that of Shock 1, this may suggest that the density of the DPE is also important during compression, even if density is not expected to play a role in dayside coupling in more steady conditions. This may be a result of stronger initial compression of the magnetosheath plasma, and hence greater reconnection electric field. However this is difficult to distinguish from other non-linear dependencies in the magnetosheath field strength and flow speed, such as any overshoots at the shock front. Note that the jump in IMF strength will also contribute to the increase in the reconnection rate, but this is consistent across all of the shocks.
	
	For Shocks 1 and 5, i.e. the due southward IMF cases, the initial peak is followed by another increase after 3 min and 2 min respectively. In both cases this timing clearly corresponds to the expansion of the magnetopause after reaching its minimum stand-off distance; it therefore appears that the motion of the boundary modulates the reconnection rate, since the inflow speed of the magnetosheath plasma changes in the rest frame of the separator. This effect is less clear for Shocks 2-4, though increases around the time of expansion are still evident. Additional smaller-scale variations are seen for all the shocks, suggesting further non-linear behaviour in the magnetosheath plasma e.g. due to waves reflecting between the bow shock and the magnetopause, as well as further oscillations along the 3-D magnetopause. Whilst Shocks 1-4 show sustained enhanced rates for the whole 5 min period shown, Shock 5 shows a sharp decline after 2 min. This indicates faster enhancement of nightside reconnection which gradually dominates the dayside rate.
	
	\subsection{Local Reconnection Rate}  \label{sec:resistive_runs}

	The variation in the reconnection rate clearly shows a complex time-dependence that would not be predicted by an empirical coupling function that takes just the upstream solar wind conditions as an input. The shorter timescale, non-linear response along the 3-D X-line may thus render these functions inaccurate for capturing dayside coupling during SI/SSC and in general for the immediate response to strong DPEs. To understand this behaviour in more detail, we now utilise our resistive runs to closely examine how the pressure enhancement generates an increased reconnection electric field. 
	
	The local reconnection rate is calculated as the electric field parallel to the separator, which we denote $\mathbf{E}_\parallel$ and calculate as $\mathbf{E}_\parallel = \eta\mathbf{J}_\parallel$. This ignores any motional electric field component ($\mathbf{E} = -\mathbf{v}\times\mathbf{B}$) which is frame-dependent and hence would otherwise have to be subtracted from the total electric field by transforming into the local inertial frame along the moving separators. This also does not capture any contributions due to numerical diffusion, but as explained earlier this is dominated by the resistive diffusion and would introduce further model-dependence in the results. 
	
	It should be stressed that the exact magnetosheath conditions depend on how the simulation captures shocks, and that we are interested only in the general trends seen. We do not compare these rates to the earlier values, since the resistivity will likely result in significant differences during compression, but rather use the resistive simulations to provide further insight into the overall behaviour. For this purpose we trace the dayside separators out in full for Shocks 1, 3 and 5: we use the same approach as used earlier for Shocks 2-4, but this time for the 180$^\circ$ clock angle cases as well.
	
\begin{figure}[h!]
		\centering
		\includegraphics[width=\linewidth]{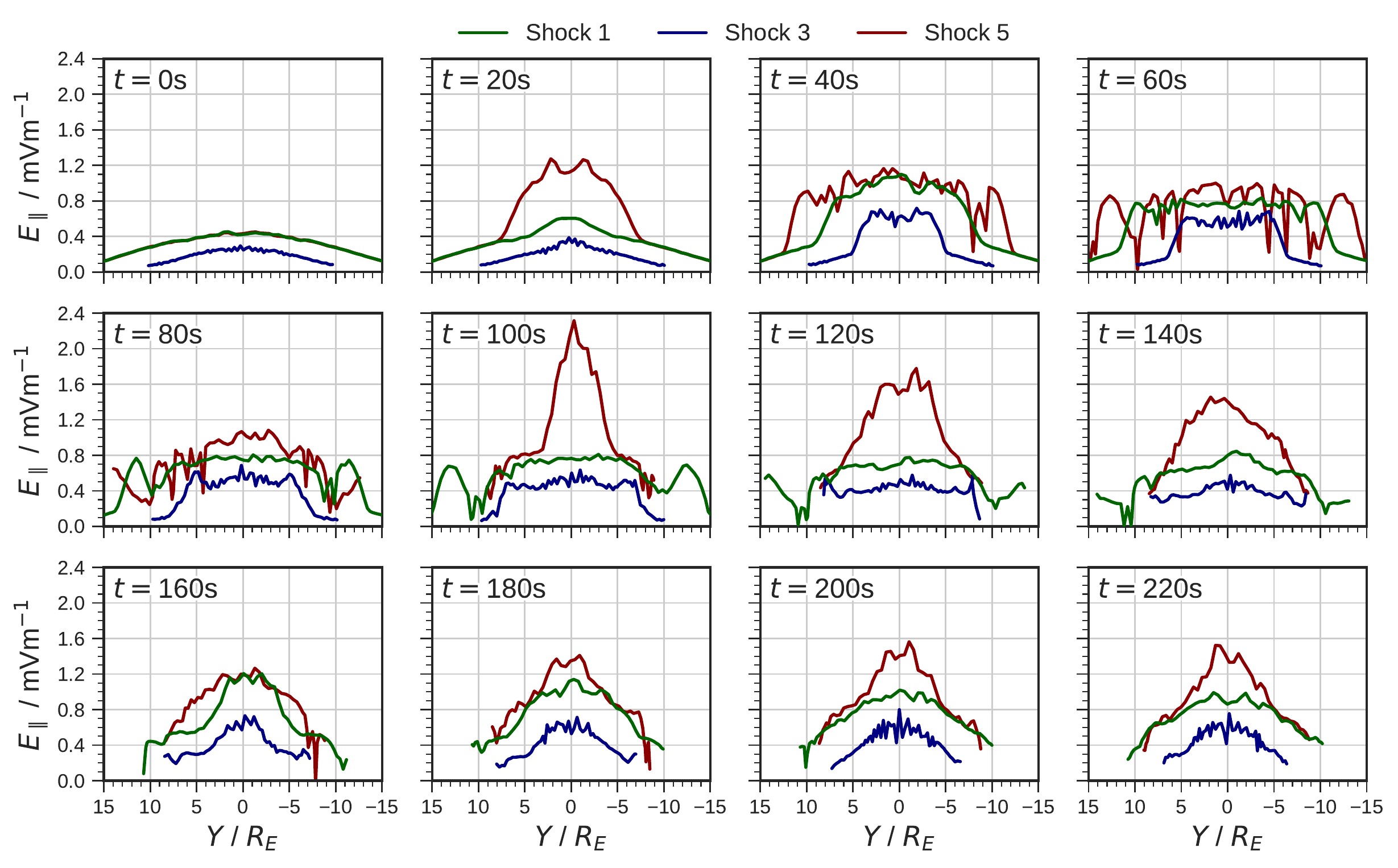}
		\caption{Local reconnection rate $E_{||} = \eta J_{||}$ along the dayside separator for Shocks 1, 3 and 5 over time, using the resistive simulations. For Shocks 1 and 3, $t = 0$ corresponds to 7340 s; for Shock 5, $t = 0$ corresponds to 7280 s.}
		\label{fig:Shocks_Epar} 
	\end{figure}

	Figure \ref{fig:Shocks_Epar} shows the reconnection rate along the separator for Shocks 1, 3 and 5 as the shock propagates over the dayside, with the time $t$ the same as in Figure \ref{fig:flux_change_vs_time}. We focus first on the case of Shock 1. At $t = 0$ the rate is just dependent on the preceding solar wind conditions, peaking around the subsolar point at $\sim0.4$ mVm$^{-1}$. After the shock reaches the magnetopause at 20 s the electric field is amplified, and the peak value at 40 s is almost twice that at 20 s. The propagation of the initial shock front is associated with a pair of local peaks in the electric field in the dawk and dusk hemispheres, due to compression of the magnetosheath magnetic field, which then travel down the flanks and past the terminator plane after 120 s. If the amplitude of these travelling peaks remains sufficiently high, this indicates a possibility for transient magnetopause reconnection on the nightside during such an event.
	
	Thus the local reconnection rate is clearly enhanced where the shock makes contact with the magnetopause, and may even trigger reconnection at locations where it would not normally be expected. Another amplification up to $\sim1.2$ mVm$^{-1}$ is then seen at 160 s, triple the intensity at 0 s, which coincides with the secondary peak in Figure \ref{fig:flux_change_vs_time} due to the expansion of the compressed magnetopause in response to waves reflected off the inner boundary. After this the reconnection rate begins to relax as the magnetopause motion is gradually arrested (as seen in Figure \ref{fig:flux_change_vs_time}). 
	
	For Shock 3 (i.e. a 90$^\circ$ clock angle) the initial state prior to the shock arrival has a weaker reconnection electric field than for purely southward IMF, and shows a smaller enhancement from 20-40 s, but still reaches over twice the peak amplitude than at 0 s. The same effect of enhanced electric field along the propagating shock front is seen up to 120 s, as well as a sudden rise at 160 s which again as for Shock 1 is roughly triple the initial intensity. Therefore the factor of increase in the local reconnection rate appears to depends solely on the increase in dynamic pressure, and the effect is not unique to the most geoeffective IMF orientations.
	
	Finally we examine Shock 5 (i.e. 180$^\circ$ and a higher dynamic pressure) for which due to its greater shock speed we begin from an earlier time window, starting at 7280 s. The later timesteps also represent a more relaxed state where the magnetopause is relatively static, with $t = 140$ s corresponding roughly to $t = 220$ s for Shocks 1 and 3 with respect to the shock position. This means that individual stages in the evolution are not directly comparable, but still demonstrate the same trends. As before the reconnection rate is enhanced at 20 s, and the shock propagation generates the same local peaks down the flank magnetopause; during this phase the peak amplitude is $\sim1.2$ mVm$^{-1}$, compared to $\sim0.8$ mVm$^{-1}$ for Shock 1. The sudden spike at the onset of magnetopause expansion reaches $\sim2.4$ mVm$^{-1}$, which is twice that seen for Shock 1, indicating that a higher dynamic pressure results in stronger shock disturbances reflected off the inner boundary.
	
	\subsection{Magnetic Null Points}
	
\begin{figure}[h!]
		\centering
		\includegraphics[width=\linewidth]{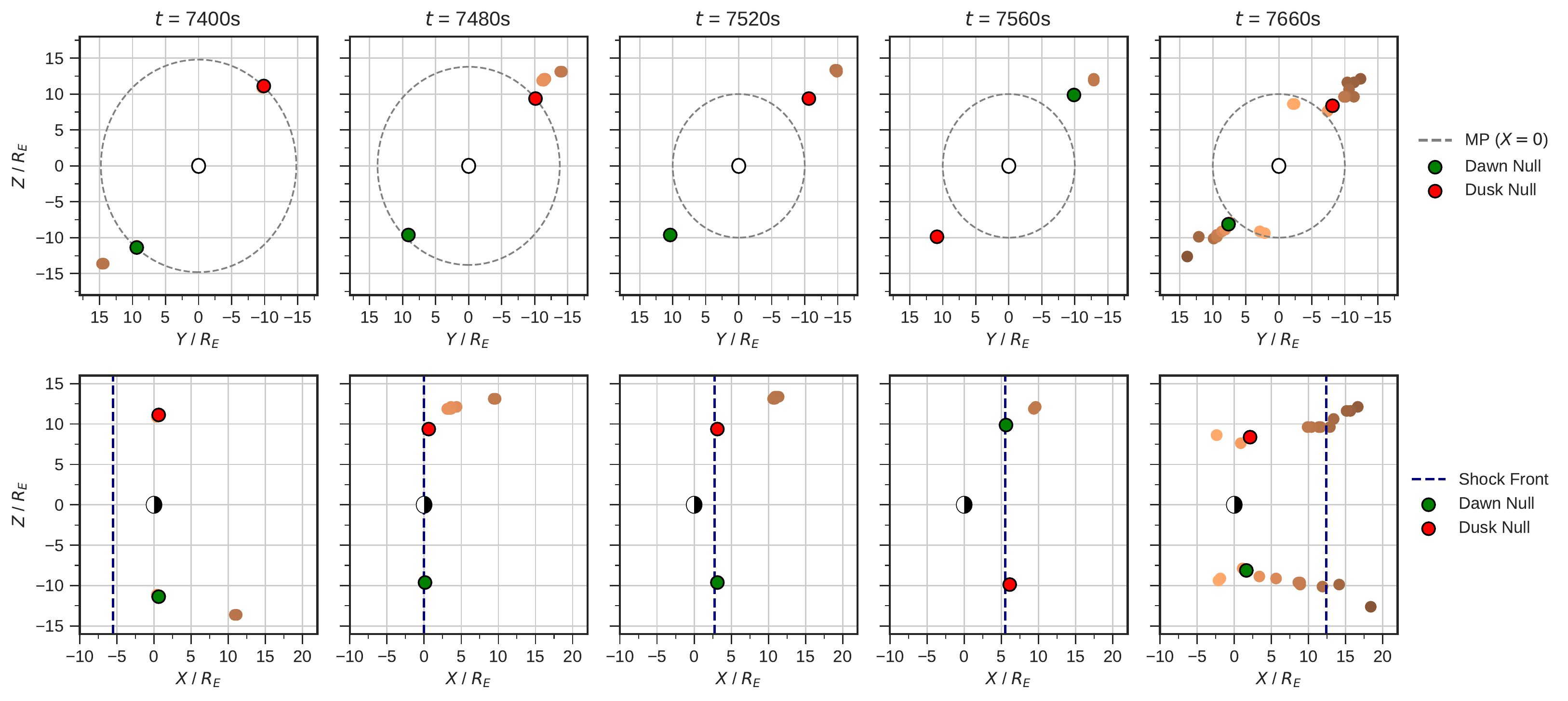}
		\caption{Magnetic null points near the magnetopause during the propagation of Shock 3, colored increasingly dark with distance downtail. The nulls at dawn and dusk closest to the vacuum superposition solutions are labelled in green and red, respectively. The intersection of the magnetopause with the terminator plane and the location of the shock front are indicated, and are approximate.}
		\label{fig:shock_3_nulls} 
	\end{figure}

	To gain more insight into the effect of the shock on the magnetopause topology as it enters the nightside, we can study the motion of magnetic null points, which as described earlier mark the end points of the magnetic separator. The nulls are found using the method described in \citeA{Eggington2020} and based on that of \citeA{Haynes2007} in which we interpolate within grid cells containing reversals in each component of $\mathbf{B}$, and we include only those along the magnetopause, i.e. ignore those within the magnetotail current sheet. We focus only on the collisionless (zero resistivity) simulation of Shock 3, for which $\theta_{IMF} = 90^\circ$, since for more southward IMF (where the field strength along null-null lines tends to zero, rendering them unstable to breakup) the nulls exist in much greater number and are harder to track individually. The evolution of the null points is shown in Figure \ref{fig:shock_3_nulls}. The `terminating' nulls are defined as those closest to vacuum predictions (using formulae from \citeA{Yeh1976}), and are coloured in red and green. The approximate location of the magnetopause at $X$ = 0 is inferred from Figure \ref{fig:pressure_shock}.
	
	Prior to the shock arrival the terminating nulls lie along the terminator plane, as expected, sitting at high latitudes at both dawn and dusk. Additional nulls are seen down the flanks, possibly associated with local instabilities, but we find these only exist in closely-situated pairs and are short-lived, and so do not contradict with the topology of a single dayside separator. By 7480 s contact with the dawn and dusk null points is made; these are then dragged anti-sunward as the shock propagates, and proceed into the nightside over the following 80 s. This must arise from the deformation of the magnetopause and the resulting change in the orientation of the magnetospheric and draped magnetosheath fields. We note that the resistive simulations do not reproduce such a large displacement of the nulls. This suggests either that a large resistivity allows diffusion of the field in such a way that the topology is better preserved, or that the null motion could be a peculiarity of the grid effects.
	
	By 7660 s new nulls have appeared near the terminator plane, closer to the vacuum predictions, though a large number of nulls still exist trailing the shock front as far as 20 $R_E$ downtail. Two pairs of nulls are seen at high latitudes around $Y\sim\pm2$ $R_E$, which may indicate high-latitude reconnection typically associated with northward IMF. This may also represent the formation of a new, steadier dayside separator that remains in place during the post-shock conditions. Whilst we have not shown the other clock angle cases here, individual nulls will still be carried downstream by the shock irrespective of IMF orientation, though the topological effect is most easily seen for the 90$^\circ$ clock angle presented.
	
	The result of this displacement is that the length of the separator connecting these nulls is extended. This does not automatically imply magnetopause reconnection is active over the portion extending into the nightside, but our finding from Figure \ref{fig:Shocks_Epar} that the transient enhancement in the reconnection rate reaches the terminator plane suggests this may be possible. Whilst we cannot be certain that this null behaviour is not a unique effect of the numerical grid, it is clear that the presence of the shock greatly alters the magnetic topology even on the nightside magnetopause.
	
	
	\section{Discussion and Conclusions}
	
	In this study we have used global MHD simulations to explore in detail the response of dayside reconnection during SI/SSC, immediately following the arrival of an interplanetary shock. The shock-induced signatures which compress the magnetosphere and propagate through the system are consistent with those in previous studies using global simulations. The evolution of the magnetic separator during the compression has been shown for a variety of different IMF clock angles, dipole tilt angles and dynamic pressure enhancements. The reconnection line responds dynamically to the distortion of the magnetopause, reducing in extent as it is compressed and moving incoherently as the shock front propagates through the dayside magnetosphere. The separator appears strongly bent at the point of contact with the shock, especially for clock angles that are not due southward. This demonstrates the highly non-linear behaviour of the dayside magnetosphere during such events.
	
	The reconnection rate is enhanced after the arrival of the shock, increasing to a sharp peak value in excess of that for the eventual post-shock solar wind conditions which we attribute to piling up of the magnetosheath plasma and magnetic field. The subsequent motion of the magnetopause and oscillations within the magnetosheath appear to modulate the reconnection rate before it eventually settles into a steady-state once the magnetopause relaxes, in agreement with theoretical suggestions by \citeA{Freeman1995}. The time-evolution of the reconnection rate shows a clear clock-angle dependence, but this is weaker than that predicted by typical coupling functions and a strong intensification is seen for all IMF orientations. This could result in a greater coupling efficiency for events with weaker $E_y$, consistent with trends seen in previous studies of DPEs (e.g. \citeA{Andreeova2011a}).   
	
	By complementing our results with resistive MHD simulations we have found that the local electric field along the magnetic separator is increased at the point of contact of the shock as it propagates, leading to an enhanced reconnection rate at regions away from the subsolar magnetopause. Our results also suggest this effect may even spread to the nightside magnetopause, and magnetic null points which mark the end of the dayside separator are indeed displaced slightly tailward as the shock passes the terminator plane. In any case, the effect would allow for reconnection at different locations on the boundary than usually expected, and should only occur briefly making it difficult to verify observationally without prior predictions from simulation studies such as this.
	
	The signatures of the enhanced dayside coupling can be seen in the ionospheric polar cap. Under southward IMF the expansion of the OCB occurs near noon shortly after the onset of SSC, and spreads out to dawn and dusk. This is closely associated with transient FACs as described in previous studies (e.g. \citeA{Fujita2003a}, \citeA{Samsonov2010}). One such system merges into the pre-existing Region 1 currents to form steady bands of FAC that match the post-shock conditions. Signatures of enhanced nightside reconnection, i.e. a contraction of the OCB, are not seen for the initial few minutes after shock impact. This is consistent with simulations by \citeA{Boudouridis2021} but in contrast to observations in the same study showing an overall contraction despite an increase in dayside reconnection, suggesting a prompt and significant nightside increase. It may be that the expansion is less detectable for certain DPEs (e.g. not specifically IP shocks) if the nightside response is stronger than on the dayside, or that simulations overestimate the initial growth in the dayside rate. This would be further complicated by any MLT-dependent time-delays in the ionospheric response, but reconciling the sequence of events in MHD models with observations is a point for future investigation.
	
	As with any simulation study there are a number of caveats to address. The dissipation mechanism responsible for reconnection in most of these runs was numerical diffusion, which will inevitably be sensitive to the grid and numerics of the model. However with the use of additional resistive MHD simulations we have reproduced the enhancement in dayside reconnection, finding the same key trends. A further issue is delineating the contributions to the initial peak in the global reconnection rate by both the pile-up of shocked magnetosheath plasma and any overshoot around the propagating shock fronts. However whilst the exact magnitude of any effects may differ depending on the simulation setup, the overall physical behaviour should remain the same. This is especially true since we are only focussing on the dayside magnetosphere, meaning any conclusions should not be too dependent on the system being initialised with steady conditions. In contrast, the nightside reconnection response may be more sensitive to the particular configuration of the magnetotail. Finally, a non-uniform ionospheric conductance may result in differences in the reflected electric field in the magnetosphere, and thus in the reconnection rate response. Future work could explore this effect, and could also compare to theoretical predictions of the local reconnection rate (e.g. \citeA{Cassak2007}) as it evolves over time.
	
	Overall the results are in stark contrast to expectations from a steady model of reconnection which assumes a linear response to changes in to upstream conditions, as in empirical solar wind coupling functions. Our simulations show that the dayside magnetosphere undergoes highly non-linear behaviour in the several minutes after the arrival of an interplanetary shock, and so attempts to use these functions when estimating the rate of change of open flux during SI/SSC may not be reliable. Recent studies have similarly shown that empirical models fail to capture the complex motion of the magnetopause during such events (\citeA{Staples2020}, \citeA{Desai2021a}). Care should therefore be taken when attempting to quantify the role of enhanced reconnection in driving geomagnetic activity shortly after onset, which we also expect to be true for other discontinuities such as dynamic pressure decreases. This has implications not only for our understanding of magnetopause reconnection, but also for space weather effects in general.
	
	\acknowledgments
	
	\noindent JWBE was funded by a UK Science and Technology Facilities Council (STFC) Studentship (ST/R504816/1). RTD, LM, JPE and JPC acknowledge funding Natural Environment Research Council (NERC) grants NE/P017347/1 (Rad-Sat) and NE/P017142/1 (SWIGS). This work used the Imperial College High Performance Computing Service (doi: 10.14469/hpc/2232).
	
	\dataavailability
	
	\noindent The simulation data used in this paper will be made publicly available on the Centre for Environmental Data Analysis (CEDA) with a DOI prior to publication.
	
	\bibliography{bib}

\end{document}